\documentclass[aps,prl,floatfix,twocolumn]{revtex4}%
\usepackage{amssymb}
\usepackage{amsmath}
\usepackage{graphicx}
\usepackage{amsfonts}%
\providecommand{\U}[1]{\protect\rule{.1in}{.1in}}
\providecommand{\U}[1]{\protect\rule{.1in}{.1in}}
\providecommand{\U}[1]{\protect\rule{.1in}{.1in}}
\providecommand{\U}[1]{\protect\rule{.1in}{.1in}}
\providecommand{\U}[1]{\protect\rule{.1in}{.1in}}
\providecommand{\U}[1]{\protect\rule{.1in}{.1in}}
\providecommand{\U}[1]{\protect\rule{.1in}{.1in}}
\providecommand{\U}[1]{\protect\rule{.1in}{.1in}}
\begin{document}
\title{Direct cavity detection of Majorana pairs}
\author{Matthieu Dartiailh$^{1}$, Takis Kontos$^{1}$, Benoit Dou\c{c}ot$^{2}$ and
Audrey Cottet$^{1}$}
\date{\today}
\affiliation{$^{1}$Laboratoire Pierre Aigrain, Ecole Normale Sup\'{e}rieure, CNRS UMR 8551,
Laboratoire associ\'{e} aux universit\'{e}s Pierre et Marie Curie et Denis
Diderot, 24, rue Lhomond, 75231 Paris Cedex 05, France}
\affiliation{$^{2}$Sorbonne Universit\'{e}s, Universit\'{e} Pierre et Marie Curie, CNRS,
LPTHE, UMR 7589, 4 place Jussieu, 75252 Paris Cedex 05}

\begin{abstract}
No experiment could directly test the particle/antiparticle duality of
Majorana fermions, so far. However, this property represents a necessary
ingredient towards the realization of topological quantum computing schemes.
Here, we show how to complete this task by using microwave techniques. The
direct coupling between a pair of overlapping Majorana bound states and the
electric field from a microwave cavity is extremely difficult to detect due to
the self-adjoint character of Majorana fermions which forbids direct energy
exchanges with the cavity. We show theoretically how this problem can be
circumvented by using photo-assisted tunneling to fermionic reservoirs. The
absence of direct microwave transition inside the Majorana pair in spite of
the light-Majorana coupling would represent a smoking gun for the Majorana
self-adjoint character.

\end{abstract}
\maketitle
\affiliation{$^{1}$Laboratoire Pierre Aigrain, Ecole Normale Sup\'{e}rieure, CNRS UMR 8551,
Laboratoire associ\'{e} aux universit\'{e}s Pierre et Marie Curie et Denis
Diderot, 24, rue Lhomond, 75231 Paris Cedex 05, France}
\affiliation{$^{2}$Sorbonne Universit\'{e}s, Universit\'{e} Pierre et Marie Curie, CNRS,
LPTHE, UMR 7589, 4 place Jussieu, 75252 Paris Cedex 05}

\begin{figure}[ptb]
\includegraphics[width=1\linewidth]{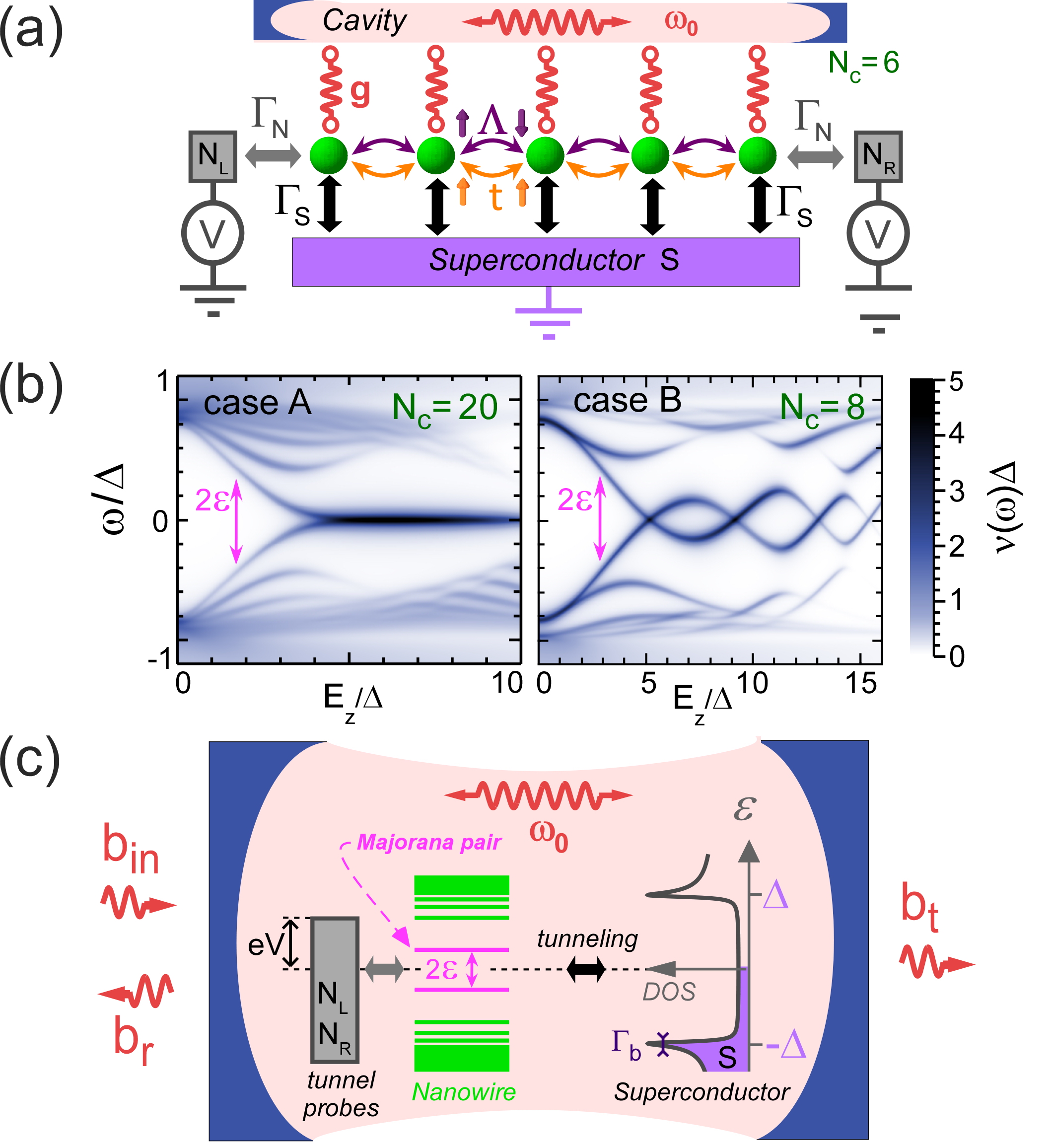}\caption{(a) Tight-binding
model of our nanocircuit with $N_{c}=6$ sites (green dots). The consecutive
sites are coupled by a tunnel hoping constant $t$ and spin-orbit terms
$\Lambda$. All sites are tunnel coupled to the superconductor $S$ with a rate
$\Gamma_{S}$. The extremal sites are tunnel coupled with a rate $\Gamma_{N}$
to normal metal contacts $N_{L(R)}$ with a bias voltage $V$. All sites are
coupled to the microwave cavity with a constant $g$. (b): DOS $\nu
(\varepsilon)$ at the ends of the chain/nanowire, versus $\omega$ and $E_{z}$,
for cases A (long nanowire with negligible coupling to $N_{L(R)}$) and B
(short nanowire and coupling to $N_{L(R)}$ and $S$ similar at zero energy).
(c): Energetic scheme of the nanocircuit placed in the microwave cavity. The
DOS in $S$ depends on the superconducting gap $\Delta$ and the broadening
parameter $\Gamma_{b}$. The cavity microwave transmission $b_{t}/b_{in}$ or
reflection $b_{r}/b_{in}$ is measured.}%
\label{setup}%
\end{figure}Majorana quasiparticles are among the most intriguing excitations
predicted in condensed matter physics\cite{Kitaev}. The Majorana particle is
equal to its own antiparticle, a property which opens up possibilities of
non-abelian statistics\cite{Ivanov} and topologically protected quantum
computation\cite{Nayak} in condensed matter systems. Low frequency conductance
measurements have prevailed so far in the experimental search for these exotic
quasiparticles. In particular, hybrid structures combining semiconducting
nanowires and superconductors have been intensively
investigated\cite{Lutchyn,Oreg}. The recent observation of zero energy
conductance peaks and pairs of peaks with an oscillatory splitting are
consistent with the existence of Majorana bound states
(MBSs)\cite{Mourik,Das,Deng,Churchill,Zhang,Albrecht}. However, it is
essential to find new tools to test more specifically the nature of these
peaks\cite{Lee,Pikulin,Rainis,Roy,Zhao,CottetQPC}.

Paradoxically, the self-adjoint character of MBSs, which draws so much
interest, also makes them very difficult to detect. Photons trapped in a high
finesse cavity are a priori very appealing for probing these elusive
excitations\cite{Wallraff}. Indeed, light, contrarily to conductance
measurements, preserves the occupation number encoded into a pair of MBSs.
Unfortunately, for the same reason, a pair of MBSs cannot exchange energy with
an electromagnetic field, which forbids any direct spectroscopy. This seems to
limit the use of light in complex
setups\cite{Trif,Schmidt,CottetMajos,Dmytruk0,Vayrynen,Virtanen}. Recently
proposed quantum computing architectures rely on an indirect MBSs-photon
coupling through a metallic Josephson
circuit\cite{Hassler,Hyart,Muller,Xue,Pekker,Ginossar,Ohm,Yavilberg,Yavilberg2}%
.

In this theoretical work, we show that the imperfections of a realistic
Majorana nanocircuit, although undesirable for quantum information
applications, can be exploited at present to characterize MBSs with microwave
techniques. First, a realistic Majorana nanocircuit must have a finite size to
remain in the coherent regime. Therefore, MBSs can have a spatial overlap.
This naturally generates, for a pair of MBSs, an energy splitting
$2\varepsilon$ and a \textit{direct} coupling $\beta$ to the cavity electric
field\cite{CottetMajos,Cottet2015}. Second, the presence of even a tiny amount
of zero-energy quasiparticles in superconducting or normal metal contacts,
which is inherent to experimental setups demonstrated so far, switches on
photo-assisted tunnel processes which involve only one partner of a Majorana
doublet. During these transitions, photons with frequency $\varepsilon$ are
exchanged between the cavity and the Majorana pair, with a rate set by $\beta
$. However, transitions at frequency $2\varepsilon$ remain forbidden
regardless of the circuit parameters. The purely longitudinal nature of
$\beta$, thereby revealed, would represent a direct signature, in the simplest
setup, of the self adjoint-character of MBSs.

Hybrid nanocircuits with superconducting parts have been coupled very recently
to microwave cavities\cite{Janvier, Bruhat,Larsen,deLange}. The light/matter
coupling in such devices can be described generically with the Hamiltonian
\begin{equation}
\hat{h}_{tot}=\hat{h}_{\mathcal{N}}+\omega_{0}\hat{a}^{\dag}\hat{a}+\hat
{h}_{\mathcal{C}}(\hat{a}+\hat{a}^{\dag})\label{htot}%
\end{equation}
with $\omega_{0}$ the cavity frequency. In a typical experiment, one
determines the cavity frequency pull $\Delta\omega_{0}$ and damping pull
$\Delta\Lambda_{0}$ from the cavity microwave response measured at frequency
$\omega_{RF}=\omega_{0}$. In case of an electric coupling scheme, from a
semiclassical linear response description, these signals correspond to
$\Delta\omega_{0}+i\Delta\Lambda_{0}=\chi(\omega_{0})$, with $\chi^{\ast}$ the
nanocircuit charge susceptibility\cite{Bruhat}. In a first approach, one can
assume that the nanocircuit spectrum is discrete. Since $\hat{h}%
_{\mathcal{N(C)}}$ are quadratic, one can use $\hat{h}_{\mathcal{N}}=%
{\textstyle\sum\nolimits_{\alpha}}
E_{\alpha}\gamma_{\alpha}^{\dag}\gamma_{\alpha}$ and $\hat{h}_{\mathcal{C}}=%
{\textstyle\sum\nolimits_{\alpha}}
M_{\alpha\beta}\hat{\gamma}_{\alpha}^{\dag}\hat{\gamma}_{\beta}+N_{\alpha
\beta}\hat{\gamma}_{\alpha}^{\dag}\hat{\gamma}_{\beta}^{\dag}+N_{\alpha\beta
}^{\dag}\hat{\gamma}_{\alpha}\hat{\gamma}_{\beta}$, with $E_{\alpha}>0$,
$\hat{\gamma}_{\alpha}^{\dag}$ Bogoliubov operators which combine electron and
hole excitations, and $\{M_{\alpha\beta},N_{\alpha\beta}\}$ matrix elements
which depend on the overlap of the cavity photonic pseudopotential with the
wavefunctions associated to $\hat{\gamma}_{\alpha(\beta)}^{\dag}%
$\cite{Cottet2015}. At zero temperature ($T=0$), this gives $\chi^{\ast
}(\omega_{0})\simeq%
{\textstyle\sum\nolimits_{\alpha\beta}}
\left\vert N_{\alpha\beta}\right\vert ^{2}(\omega_{0}-E_{\alpha}-E_{\beta
}+i0^{+})^{-1}/2$. Importantly, due to the Pauli exclusion principle, one has
$N_{\alpha\alpha}=0$. Hence, $\chi(\omega_{0})$ does not involve transitions
between electron and holes associated to conjugated operators $\hat{\gamma
}_{\alpha}^{\dag}$ and $\hat{\gamma}_{\alpha}$. This selection rule can be
extended to $T\neq0$ or a level broadening smaller than the inter-level
separation\cite{SM}. However, having a nanocircuit response at $\omega
_{0}=2E_{\alpha}$ is possible provided there exists a state degeneracy
$E_{\alpha}=E_{\alpha^{\prime}}$ in the nanocircuit\cite{SM}, as observed in
spin-degenerate superconducting atomic contacts\cite{Janvier,Skoldberg}. In
contrast, lifting degeneracies is crucial to obtain MBSs. In elementary
models\cite{Lutchyn,Oreg}, when a finite portion of nanoconductor is driven to
its topological phase, one non-degenerate Bogoliubov doublet ($\hat{\gamma
}_{1}^{\dag},\hat{\gamma}_{1}$) approaches alone the zero energy area to form
a pair of MBSs described by self-adjoint creation operators $\hat{m}_{L}%
=(\hat{\gamma}_{1}^{\dag}+\hat{\gamma}_{1})/\sqrt{2}$ and $\hat{m}_{R}%
=i(\hat{\gamma}_{1}^{\dag}-\hat{\gamma}_{1})/\sqrt{2}$ such that one has, at
low energies, $\hat{h}_{\mathcal{N}}=\varepsilon\hat{\gamma}_{1}^{\dag}%
\hat{\gamma}_{1}=i\varepsilon\hat{m}_{L}\hat{m}_{R}+(\varepsilon/2)$ with
$\varepsilon=E_{11}$. An important signature of this scenario is the absence
of direct microwave transitions in the Majorana subspace, i.e. $\chi
(\omega_{0}=2\varepsilon)\simeq0$ due to $N_{11}=0$. Remarkably, this occurs
even when the Majorana-cavity coupling is finite, i.e. $\hat{h}_{\mathcal{C}}$
contains a term in $i\beta\hat{m}_{L}\hat{m}_{R}$ with $\beta=M_{11}$. Using a
spin analogy, the cavity and the MBSs can only have a longitudinal coupling,
which is not able to change the state of the Majorana pair, i.e. $\hat
{h}_{\mathcal{C}}$ and $\hat{h}_{\mathcal{N}}$ are represented by collinear
vectors in the Bloch sphere associated to the Majorana subspace. This gives a
smoking gun for the non-degenerate $(\hat{\gamma}_{1},\hat{\gamma}_{1}^{\dag
})$ electron-hole conjugated pair, or equivalently, the pair of self-adjoint
excitations ($\hat{m}_{L}$, $\hat{m}_{R}$). Importantly, the absence of direct
microwave transitions in the Majorana doublet is meaningful only if one can
confirm $\beta\neq0$. Furthermore, the absence of direct transitions should be
robust when the control parameters of the nanocircuit are varied, to discard
any accidental cancellation of $\chi(\omega_{0})$. We use below a specific
example to show how these tasks can be achieved.\begin{figure}[ptb]
\includegraphics[width=1\linewidth]{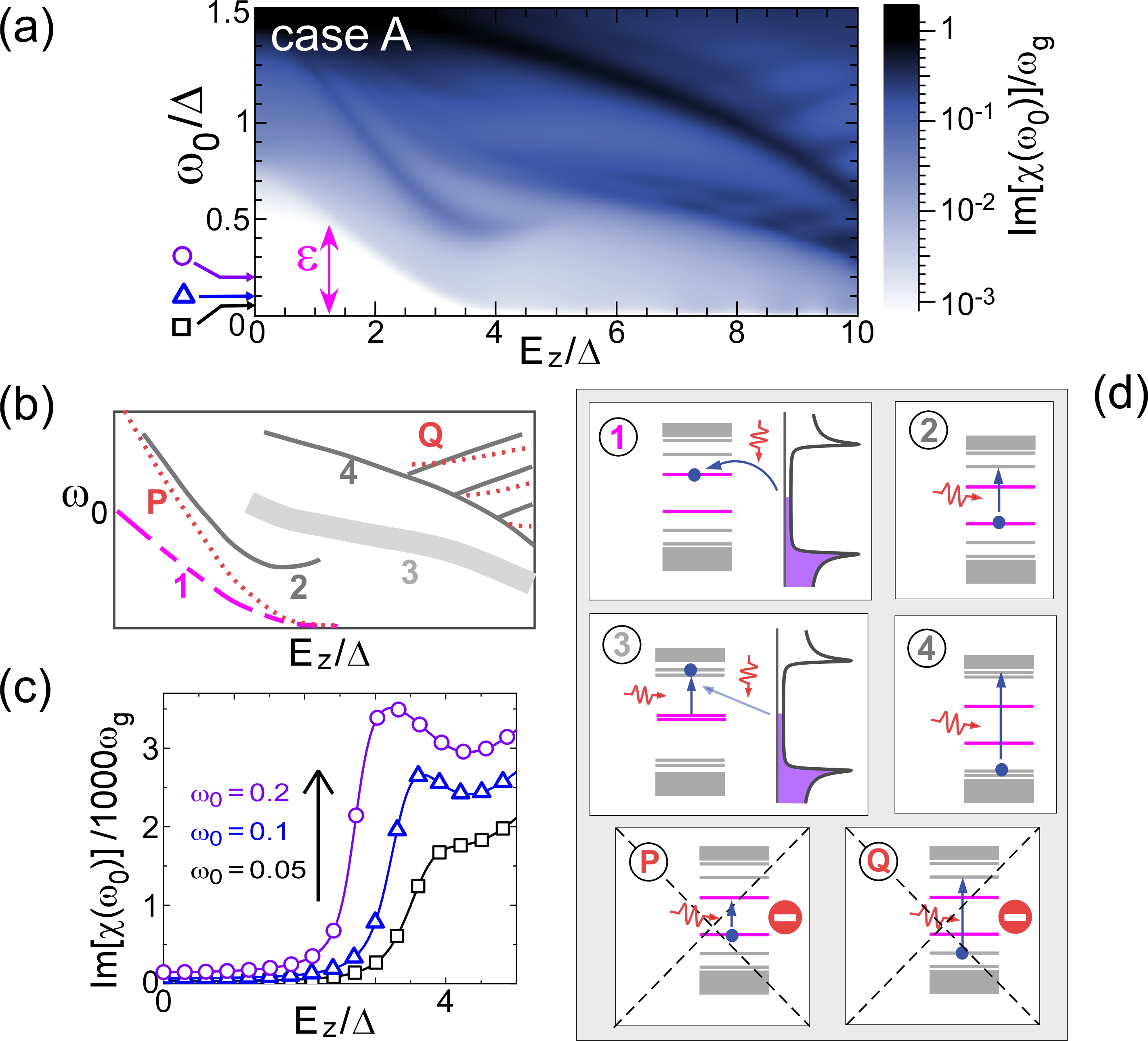}\caption{(a):
$\operatorname{Im}[\chi(\omega_{0})]$ versus $\omega_{0}$ and $E_{z}$ for case
A with $V=0$ (b) Scheme of the main features appearing panel (a). (c):
$\operatorname{Im}[\chi(\omega_{0})]$ versus $E_{z}$ for constant values of
$\omega_{0}$ corresponding to the symbols in panel (a). (d): Processes
contributing to the features of panel (b), and forbidden processes P and Q.}%
\label{Figure2}%
\end{figure}

We now consider a semiconducting nanowire subject to spin-orbit coupling and a
Zeeman field. The nanowire is tunnel coupled along its whole length to a
superconducting contact $S$. It is also coupled at both ends to normal metal
contacts $N_{L}$ and $N_{R}$ which enable the measurement of the nanowire
density of states $\nu(\omega)$. We describe this circuit with a
phenomenological one-dimensional tight-binding chain $\hat{h}_{\mathcal{W}}=%
{\textstyle\sum\nolimits_{n}}
[\hat{d}_{n}^{\dag}\left(  E_{z}\hat{\sigma}_{z}-\mu\hat{\sigma}_{0}\right)
\hat{d}_{n}-(\hat{d}_{n}^{\dag}(t\hat{\sigma}_{0}+\Lambda\hat{\sigma}_{y}%
)\hat{d}_{n+1}+h.c.)]$ with $\hat{d}_{n}^{\dag}=\{d_{n\uparrow}^{\dag
},d_{n\downarrow}^{\dag}\}$ and $d_{n\sigma}^{\dag}$ the creation operator for
an electron with spin $\sigma$ in site $n\in\lbrack1,N_{c}]$ of the chain
(Fig.\ref{setup}a). We denote by $E_{z}$ the Zeeman field on the sites, $\mu$
the sites chemical potential, which can be tuned with a gate voltage, $t$ the
hoping constant between the sites and $\Lambda$ the spin-orbit constant. So
far, studies on MBSs coupled to cavities have reduced the effect of $S$ to an
effective pairing term added to $\hat{h}_{\mathcal{W}}$%
\cite{Trif,Schmidt,CottetMajos,Dmytruk0}. However, a realistic model must also
take into account level broadening and dissipation. Therefore, we describe
explicitly tunneling to $S$ and $N_{L(R)}$ with an Hamiltonian $\hat
{h}_{\mathcal{R}}$ (see \cite{SM} for details). This leads us to introduce the
tunnel rate $\Gamma_{N}$ between site $1(N_{c}\mathbf{)}$ and contact
$N_{L(R)}$ and the tunnel rate $\Gamma_{S}$ between site $n\in\lbrack1,N_{c}]$
and contact $S$. The DOS of $S$ depends on the superconducting gap $\Delta$
but also on a phenomenological parameter $\Gamma_{b}$ which accounts for a
broadening of the BCS peaks and a finite low energy DOS (Fig.\ref{setup}c).
Such effects can be caused by a finite magnetic field\cite{LeeSilvano,remark}.
We consider short chains and choose parameters such that few subgap levels are
visible in the DOS of the nanowire, like in recent
experiments\cite{Zhang,Albrecht}. More precisely, for case A, one has
$N_{c}=20$, $t=2.5\Delta$, $\Lambda=5\Delta$, $\Gamma_{b}=0.1\Delta$,
$\Gamma_{N}=0.001\Delta$, $\mu=6\Delta$ and for case B one has $N_{c}=8$,
$t=5\Delta$, $\Lambda=4\Delta$, $\Gamma_{b}=0.05\Delta$, $\Gamma_{N}%
=0.2\Delta$, $\mu=5.5\Delta$. We use $E_{z}=\Delta$, $\Gamma_{S}=5.5\Delta$
and $k_{B}T=0.01\Delta$ for both cases. The density of states $\nu(\omega)$ at
the ends of the nanowire reveals the occurrence of a pair of MBSs above a
critical Zeeman field (Fig. \ref{setup}b). These MBSs show an oscillatory
energy splitting with $E_{z}$ and $\mu$ for a very short chain (case B), or
stick to zero energy for a longer chain (case A)\cite{Stanescu}.

We assume that the above nanocircuit is embedded in a microwave cavity. Hence,
we use Hamiltonian $\hat{h}_{tot}$ of Eq.(\ref{htot}) with $\hat
{h}_{\mathcal{N}}=\hat{h}_{\mathcal{W}}+\hat{h}_{\mathcal{R}}$ and $\hat
{h}_{\mathcal{C}}=g\sum\nolimits_{n}\hat{d}_{n}^{\dag}\hat{d}_{n}$. This last
term means that cavity photons modulate the chemical potential of site $n$
with a coupling constant $g$\cite{Cottet2015}. To treat on the same footing
internal nanowire transitions and tunneling to the reservoirs, we use a
Keldysh approach\cite{Skoldberg,Schiro,Agarwalla,Agarwalla2,Bruhat}. We use
Python and Numba, a LLVM-based Python compiler\cite{Numba}, to calculate
numerically $\Delta\omega_{0}+i\Delta\Lambda_{0}=\chi(\omega_{0})$
with\cite{Bruhat}
\begin{equation}
\chi^{\ast}(\omega_{0})=-ig^{2}%
{\textstyle\int}
\frac{d\omega}{4\pi}\mathrm{Tr}\left[  \mathcal{\check{S}}(\omega
)\mathcal{\check{G}}^{r}(\omega)\check{\Sigma}^{<}(\omega)\mathcal{\check{G}%
}^{a}(\omega)\right]  \label{Xitot}%
\end{equation}
and $\mathcal{\check{S}}(\omega)=\check{\tau}\left(  \mathcal{\check{G}}%
^{r}(\omega+\omega_{0})+\mathcal{\check{G}}^{a}(\omega-\omega_{0})\right)
\check{\tau}$. Above, the retarded and advanced multisite Green's functions
$\mathcal{\check{G}}^{r/a}$ and the lesser self energy $\check{\Sigma}%
^{<}(\omega)$ can be calculated from $\hat{h}_{\mathcal{N}}$, while
$\check{\tau}$ takes into account the structure of the photon/particle
coupling in the Nambu$\otimes$spin space \cite{SM}.

We focus on the dissipative response $\Delta\Lambda_{0}=\operatorname{Im}%
[\chi(\omega_{0})]$ of the cavity, which should naturally reveal the effects
of dissipative reservoirs. We first consider case A where $\varepsilon
\rightarrow0$ in the topological phase of the nanowire. The $\omega_{0}-E_{z}$
map of $\operatorname{Im}[\chi(\omega_{0})]$, shown in Fig.\ref{Figure2}a with
$\omega_{g}=g^{2}/\Delta$, reveals a wealth of features, sketched in
Fig.\ref{Figure2}b. Feature \textcircled{\tiny{1}} is shown in more details in
Fig.\ref{Figure2}c for constant values of $\omega_{0}$. It consists of a step
at $\omega_{0}=\varepsilon$, which is the energy distance between one MBS and
the Fermi level of the reservoirs. For case A, the effects of the $N_{L(R)}$
reservoirs on $\chi$ can be disregarded due to the vanishing $\Gamma_{N}$.
Therefore, feature \textcircled{\tiny{1}} can be attributed to photo-induced
tunneling between the MBSs and the residual subgap DOS of $S$, as represented
in Fig. \ref{Figure2}d\textcircled{\tiny{1}}. In practice, it is possible to
have a well-grounded $S$ contact by realizing a direct connection between $S$
and the cavity ground plane\cite{ground,Bruhat}. In this case, feature
\textcircled{\tiny{1}} can exist only if the MBSs are directly coupled to
cavity photons, i.e. $\beta\neq0$. In spite of this coupling, no transition
occurs at $\omega_{0}=2\varepsilon$ (red dotted line in Fig.\ref{Figure2}b).
The simultaneous presence/absence of a step/resonance at $\varepsilon
$($2\varepsilon$) occurs on a wide range of $E_{z}$. We have therefore
obtained a signature of the Majorana self-adjoint character\cite{decoherence}.
\begin{figure}[ptb]
\includegraphics[width=1\linewidth]{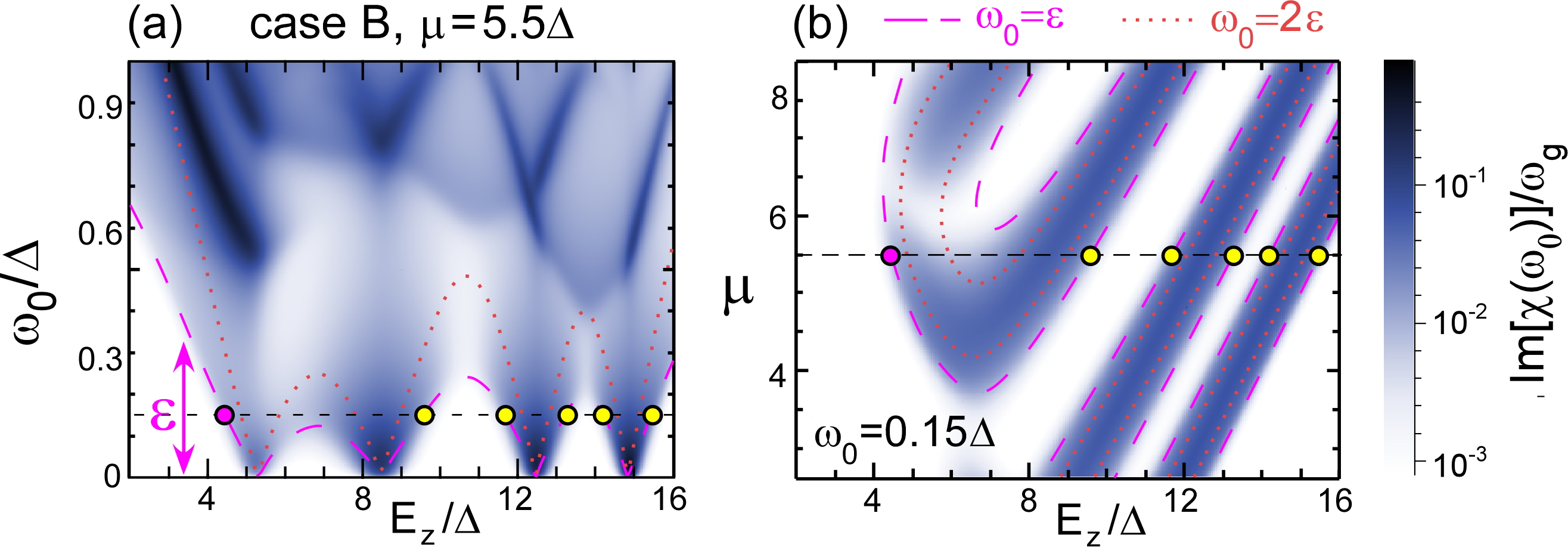}\caption{(a):
$\operatorname{Im}[\chi(\omega_{0})]$ versus $\omega_{0}$ and $E_{z}$ for case
B and $V=0$ (b): Corrresponding $E_{z}-\mu$ map for $\omega_{0}=0.15\Delta$.
The pink circle corresponds to feature \textcircled{\tiny{1}} of
Fig.\ref{Figure2} while the features with the yellow circles are specific to
the short nanowire case.}%
\label{Figure3}%
\end{figure}More precisely, these features indicate that we are in presence of
a non-degenerate electron/hole conjugated pair, which is the natural precusor
of a Majorana pair. It is then important to check from $\nu(\omega)$ that
$\varepsilon$ vanishes with $E_{z}$ (or shows several zero-energy crossings),
as a signature of the spatial isolation of the two MBSs formed out of the
non-degenerate electron/hole pair.

We now demonstrate the robustness of our main results to variations of the
nanowire spectrum. Figure \ref{Figure3}.a shows the $\omega_{0}-E_{z}$ map of
$\operatorname{Im}[\chi(\omega_{0})]$ in the case of a shorter nanowire (case
B). Feature \textcircled{\tiny{1}} persists in this limit, as indicated by the
pink circle. Besides, the yellow circles indicate new steps caused again by
photo-assisted tunneling from/to the MBSs at $\omega_{0}=\varepsilon$. These
steps can have a contrast significantly stronger than feature 1. Meanwhile, no
resonant feature is visible along the $\omega_{0}=2\varepsilon$ contour
indicated by the red dotted line. This extends the possibilities for testing
the longitudinal character of the coupling between the Majorana doublet and
the cavity. Importantly, feature 1 of case A has an amplitude of
$3~10^{-3}g^{2}/\Delta$. With $\Delta=180~\mathrm{\mu eV}$ and a site-cavity
coupling $g=2~\mathrm{\mu eV}$, this corresponds to $15~\mathrm{kHz}$. The
pink circle features of case B have an amplitude of $6.3~10^{-2}g^{2}/\Delta$
which corresponds to $340~\mathrm{kHz}$. These signals are small but within
experimental reach\cite{Stehlik2,Samkharadze}, although the residual zero
energy DOS in $S$ is small, i.e. $r\sim5\%[2.5\%]$ in case A[B], which leads
to the ''hard gap'' situation similar to Ref.\cite{HardG}. Note that in case
B, tunneling quasiparticles can be provided by both the $S$ and $N_{L(R)}$
reservoirs, which contribute similarly to the broadening of the low-energy
MBSs due to our choice of parameters. We will see later how to distinguish
these contributions thanks to a finite bias voltage. Noticeably, an isolated
zero energy crossing of two ordinary Andreev bound states could be caused by a
trivial spin-degeneracy lifting. However, in this case, the frequency of
feature \textcircled{\tiny{2}}, which corresponds to an internal nanowire
transition inwards/outwards the pair, will not depend on $E_{z}$, in contrast
with the Majorana case where it strongly depends on $E_{z}$ due to the
topological phase transition\cite{SM}. Therefore, our setup is also able to
rule out the case of a trivial superconducting wire with time reversal
symmetry breaking impurities.

In practice, to measure experimentally signals similar to Figs.\ref{Figure2}.a
and \ref{Figure3}a, one must vary $\omega_{0}$. In principle, this is
technically possible\cite{Palacios,Sandberg}. However, it is useful to adapt
our predictions for standard setups with a fixed $\omega_{0}$. In this case,
other parameters must be changed to characterize the
nanocircuit\cite{Viennot2}. In Fig.\ref{Figure3}b, we show $\Delta\Lambda_{0}$
versus $E_{z}$ and $\mu$, for case A. We use $\omega_{0}=0.15\Delta$ which
corresponds, with the gap $\Delta=180~\mathrm{\mu eV}$ of Al, to the value
$\omega_{0}=6.6~\mathrm{GHz}$ compatible with present microwave techniques. In
these conditions, $\Delta\Lambda_{0}$ shows an ensemble of photo-assisted
tunneling stripes which reveal the well known oscillations of $\varepsilon$
with $E_{z}$ and $\mu$. The correspondence between Figs. \ref{Figure3}a and b
is given by the pink and yellow circles. The stripes are absent for low values
of $E_{z}$, where the nanowire makes the transition to its non-topological
phase and the MBSs thus disappear. The $\omega_{0}=2\varepsilon$ contours are
shown with red dotted lines in Fig.\ref{Figure3}b. They do not correspond to
any remarkable feature in the $\mu-E_{z}$ map of $\operatorname{Im}%
[\chi(\omega_{0})]$, contrarily to the $\omega_{0}=\varepsilon$ contours (pink
dashed lines). Importantly, in an experiment, the red and pink contours can be
determined independently from any theory, by performing a conductance
measurement on $N_{L(R)}$ to get $\nu(\omega)$. We conclude that in the case
where $\omega_{0}$ cannot be varied, a $\mu-E_{z}$ map of $\Delta\Lambda_{0}$
and $\nu(\omega)$ gives an efficient way to characterize the light-matter
coupling in our circuit. \begin{figure}[ptb]
\includegraphics[width=1\linewidth]{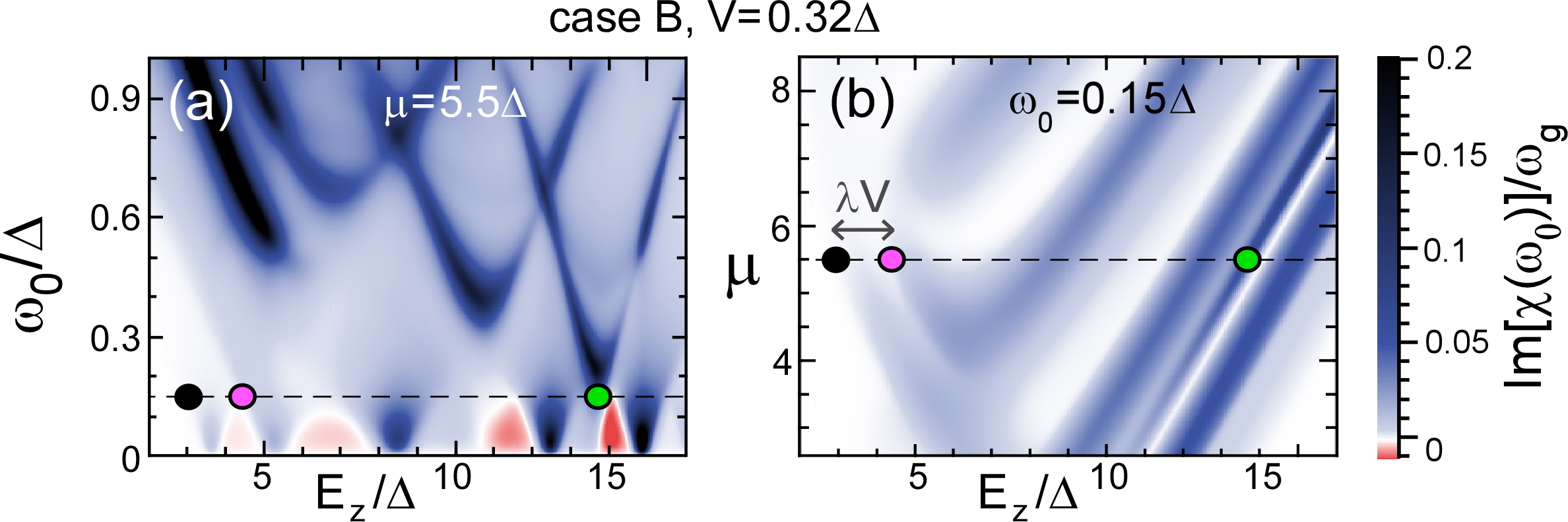}\caption{(a):
$\operatorname{Im}[\chi(\omega_{0})]$ versus $\omega_{0}$ and $E_{z}$ for case
B and $eV=0.32\Delta$ (b) Corresponding $\mu-E_{z}$ map of $\operatorname{Im}%
[\chi(\omega_{0})]$ for $\omega_{0}=0.15\Delta$}%
\label{Figure4}%
\end{figure}

We show below that applying a bias voltage to the hybrid nanocircuit-cavity
system\cite{Liu2,Viennot,Stockklauser} enables one to discriminate processes
involving the different fermionic reservoirs and to further check that the
MBSs are well coupled to cavity photons. Fig.\ref{Figure4}a shows the
$\omega_{0}-E_{z}$ map of $\operatorname{Im}[\chi(\omega_{0})]$ for case B,
with a finite bias voltage $V$ applied simultaneously to $N_{L}$ and $N_{R}$.
We observe clear differences with the case $V=0$ of Fig.\ref{Figure3}a. First,
a new step marked by the black circle appears, due to tunneling between the
MBSs and the $N_{L(R)}$ reservoirs, at $\omega_{0}=\varepsilon-eV$. Meanwhile,
the step marked with the pink circle at $\omega_{0}=\varepsilon$ persists and
is now only due to tunneling to $S$.\ The separation $\lambda V$ between the
black and pink circles, which appears for a finite $V$, is also well visible
in the $\mu-E_{z}$ map of $\operatorname{Im}[\chi(\omega_{0})]$ (see
Fig.\ref{Figure4}b where $\lambda\simeq e/\Delta(\partial\varepsilon/\partial
E_{z})$). Second, photon emission revealed by $\operatorname{Im}[\chi
(\omega_{0})]>0$ appears for $V>\varepsilon+\omega_{0}$, due to inelastic
tunneling between $N_{L(R)}$ and the upper MBS (red areas in Fig.\ref{Figure3}%
a). We conclude that the use of a bias voltage enables a differentiation of
the processes involving the $N_{L(R)}$ and $S$ contacts. The persistence of
the pink circle feature ensures that cavity photons modulate the potential
difference between the MBSs and $S$. Interestingly, for $V>\varepsilon$, the
upper MBS becomes populated so that internal transitions to upper Andreev
levels appear at remarkably low frequencies (see e.g. green circle in
Figs.\ref{Figure4}c and d). This represents another signature of the
photon-MBS coupling, although it is not the coupling constant $\beta=M_{11}$
which is involved in this case but rather $N_{1\alpha}$ with $\alpha\neq1$ and
$E_{\alpha}>\varepsilon$.

In any detection setup, false positive detection events can occur. In our
case, a false positive detection of MBSs could happen in the (unlikely) case
of a pair of extended non-degenerate Andreev bound states which would have
accidentally an energy splitting with the same magnetic field dependence as
the non-local pair of localized Majorana modes of Fig. \ref{setup}b, and a
non-conclusive feature 2. To rule out such a situation , one could perform
supplementary tests readily accessible in our setup, such as non-local
transport measurements using the $S$, $N_{L}$ and $N_{R}$ contacts, with, for
instance, a varying pair splitting $2\varepsilon$ (see for instance
Refs.\cite{Bolech:2007,Nilsson:2008,Liu:2013}).

To conclude, we have shown how to exploit photo-induced tunneling to check
that a pair of MBSs is directly coupled to cavity photons. However, the direct
microwave transitions inside the Majorana subspace remains forbidden in a wide
range of parameters. This provides a means to check the self-adjoint character
of MBSs. Importantly, this protocol is independent from any theory if the
conductance of the nanowire is measured simultaneously with the cavity
response to determine $\nu(\omega)$. Such crossed measurements are routinely
achieved with mesoscopic QED
devices\cite{Viennot,Bruhat,Frey1,Delbecq1,Petersson,Petersson2,Delbecq2,Liu,Stockklauser}%
. Our proposal relies on a nanocircuit geometry widely realized
experimentally, and which has reproducibly revealed low energy conductance
peaks. Furthermore, nanoconductors with superconducting contacts have been
coupled to microwave cavities recently\cite{Bruhat,Larsen,deLange}. Therefore,
our proposal can be straightforwardly implemented with present experimental means.

\begin{acknowledgments}
\textit{Acknowledgements: We thank M.M. Desjardins and L.C. Contamin for
useful discussions. This work was financed by the ERC Starting grant CirQys.}
\end{acknowledgments}


\begin{thebibliography}{99}
\bibitem {remark}For simplicity, we use a $\Gamma_{b}$ which is constant with
$E_{z}$.
\bibitem {ground}In practice, the coupling between the fermionic reservoirs
and the cavity photons can be characterized by tuning the nanocircuit in the
adiabatic regime where tunneling is much faster than $\omega_{0}$, and using a
strong semiclassical microwave excitation, along the lines of
Ref.\cite{Delbecq2}.
\bibitem {decoherence}Decoherence could reduce the visibility of an internal
transition between two nanowire states. However, the width of feature
\textcircled{\tiny{1}} provides an upper bound for the MBSs broadening . If
feature \textcircled{\tiny{1}} is observable and mediated by the coupling
constant $\beta$, it implies that the internal transition at $\omega
_{0}=2\varepsilon$ will also be visible in the absence of the self-adjoint property.
\end{thebibliography}
\end{document}